# Information loss, entropy production and time reversal for dissipative and diffuse processes

P. Burgholzer[1,2]

[1] *Christian Doppler Laboratory for Photoacoustic Imaging and Laser Ultrasonics*
[2] *Research Center for Non Destructive Testing (RECENDT)*
*E-mail: peter.burgholzer@recendt.at*
*Austria*

**Contents**

## 1. Introduction

In non-destructive imaging the information about the spatial pattern of a samples interior has to be transferred to the sample surface by certain waves, e.g. ultrasound or electromagnetic waves. At the sample surface these waves can be detected and the interior structure is reconstructed from the measured signals (Fig. 1). The amount of information about the interior of the sample, which can be gained from the detected waves on the sample surface, is essentially influenced by the propagation from its excitation to the surface. Scattering, dissipation, or diffusion causes entropy production and a loss of information for the propagating waves, and therefore results in a loss of resolution for imaging the interior structure. A unifying framework for treating diverse diffusion-related periodic phenomena under the global mathematical label of diffusion-wave fields has been developed by Mandelis[1], like thermal waves, charge-carrier-density waves, diffuse-photon-density waves, but also modulated eddy currents, neutron waves, or harmonic mass-transport diffusion waves.

There have been made several attempts to compensate these diffusive or dissipative effects to get a higher resolution for the reconstructed images of the samples interior. In this work it is shown that thermodynamical fluctuations limit this compensation and therefore also the spatial resolution for non-destructive imaging at a certain depth is limited.

In this publication we describe one example for diffusion and another one for dissipation and the loss of information is modeled by stochastic processes. A first example described in section 3 is thermal diffusion with temperature as a random variable. Thermography uses the time dependent diffusion of heat (either pulsed or modulated periodically) which goes along with entropy production and a loss of information. Thermal waves are a good example as they exist only because of thermal diffusion. The amplitude of a thermal wave decreases more than 500 times at a distance of just one wavelength[2].

A second example described in section 4 is photoacoustic imaging taking acoustic attenuation of the generated ultrasound wave into account. Here the pressure of the acoustic wave is the random variable and its temporal evolution is described as a stochastic process. As for any other dissipative process, the energy in attenuated acoustic waves is not lost but is transferred to heat, which can be described in thermodynamics by an entropy increase. This increase in entropy is equal to a loss of information, as defined by Claude E. Shannon[3], and no compensation algorithm can compensate this loss of information. This is a limit given by the second law of thermodynamics.

For both examples the thermodynamic entropy production is equal to the loss of information, which results in a theoretical limit for the achievable spatial resolution in the reconstructed image.

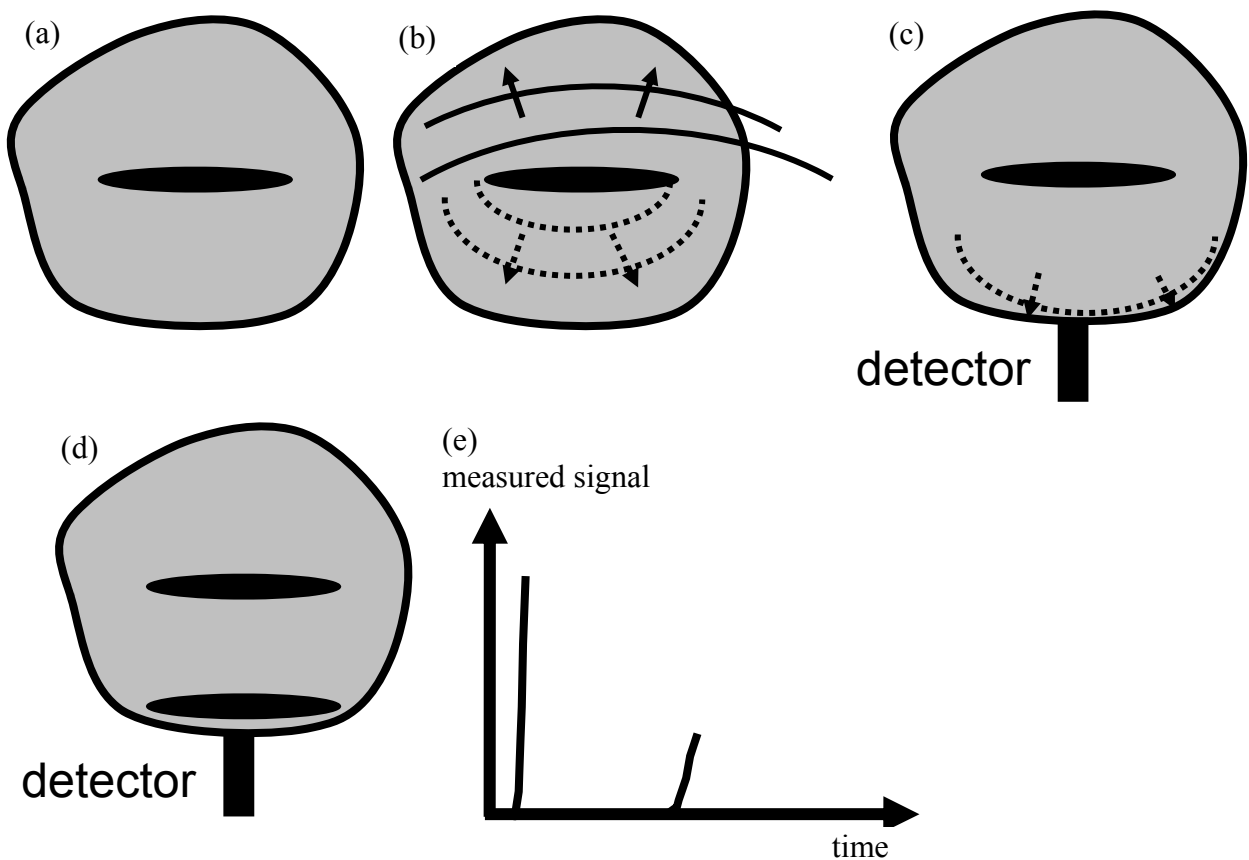

*Fig. 1. Information loss in non-destructive imaging: the spatial resolution is essentially influenced by the propagation of certain waves, e.g. ultrasound or electromagnetic waves to the surface. Scattering, dissipation, or diffusion causes entropy production and a loss of information of the propagating waves. (a) Sample containing internal structure, which should be imaged. (b) Propagation of the waves to the sample surface: entropy production goes along with fluctuations ("fluctuation-dissipation theorem", see text) and a loss of information. (c) Detection of wave at the sample surface. (d) The same structure is contained twice in the sample: just beneath the surface and at a higher depth. (e) Measured signal at the detector surface as a function of time for the sample shown in (d): due to diffusion or dissipation the signal from the deeper structure has not only smaller amplitude but it is also broadened compared to the signal from the structure just beneath the surface.*

The outstanding role of entropy and information in statistical mechanics was published in 1963 by E. T. Jaynes[4]. Already in 1957 he gave an information theoretical derivation of *equilibrium thermodynamics* showing that under all possible probability distributions with particular expectation values (equal to the macroscopic values like energy or pressure) the distribution which maximizes the Shannon information is realized in thermodynamics[5]. Jaynes explicitly showed for the canonical distribution, which is the thermal equilibrium distribution for a given mean value of the energy, that the Shannon or Gibbs entropy change is equal to the dissipated heat divided by the temperature, which is the entropy as defined in phenomenological thermodynamics[5,6]. This "experimental entropy" in conventional thermodynamics is only defined for equilibrium states. By using the equality to Shannon information Jaynes recognized, that this "gives a generalized definition of entropy applicable to arbitrary nonequilibrium states, which still has the property that it can only increase in a reproducible experiment"[6].

In non-destructive imaging, the sample is not in equilibrium, e. g. for thermography or photoacoustic imaging a short pulse from a laser or a flash light induces a non-equilibrium state which in the end evolves into an equilibrium state. For states *near thermal equilibrium in the linear regime* we use in section 2.1 the theory of non-equilibrium thermodynamics presented by S.R. de Groot and P. Mazur[7]. In this regime microscopic time reversibility entails, as Onsager has first shown, a relation between fluctuation and dissipation, since one cannot distinguish between the average regression following an external perturbation or an equilibrium fluctuation[8]. This relation is known as the fluctuation-dissipation theorem and is due to Callen[9], Welton[10] and Greene[11]. It represents in fact a generalization of the famous Johnson[12] Nyquist[13] formula in the theory of electric noise. In this publication the fluctuation-dissipation theorem provides a relation between the entropy production caused by heat diffusion or acoustic attenuation and the fluctuations of the random variable around its mean value. Dissipative and diffuse processes cause entropy production and fluctuations, which can be sketched as two sides of the same coin.

Over the last two decades, time reversibility of deterministic or stochastic dynamics has been shown to imply relations between fluctuation and dissipation in systems *far from equilibrium*, taking the form of a number of intriguing equalities, the fluctuation theorem[14,15], the Jarzynski equality[16], and the Crooks relation[17]. The conceptual framework of "stochastic thermodynamics" relates applied or extracted work, exchanged heat and changes in internal energy along a single fluctuating trajectory[18]. In an ensemble one gets probability distributions for these quantities. Since dissipated heat is typically associated with changes in entropy one gets also a distribution for the entropy. In section 2.2 will use a consequence of these equalities: that the thermodynamic entropy production is equal to the relative entropy, which is a quantitative measure of the information loss.

In this work the measured temperature or pressure signal is treated as a time-dependent random variable with a mean value and a variance as a function of time. More about random variables and stochastic processes can be found e.g. in the book about Statistical Physics from J. Honerkamp[19]. An introduction to stochastic processes on an elementary level has been published by D. S. Lemons[20], also containing "On the Theory of Brownian Motion" by Paul Langevin[21]. An introduction to Markov Processes on a slightly more advanced level is given by D. T. Gillespie[22].

Most of the first part of this work has been accepted for publication (see Burgholzer and Hendorfer[23]), but printing will take some time. In the meantime the section about the Wiener process (section 2.1.1), the reconstruction in ω-space (section 3.4) and a numerical example for comparing the reconstruction in k-space and ω-space (section 3.5) have been added. We have published first results to describe an attenuated acoustic wave as a stochastic process for photoacoustic imaging in an Intech-chapter[24] (section 4). Section 5 concludes this paper and gives a comprehensive outlook on planned future work.

## 2. Information loss and entropy production

### 2.1. Gauss-Markov Processes and Entropy

In this section we shall first assume that the time varying stochastic processes will have Gauss-Markov character. In doing so we do not wish to assume that all macroscopic processes considered belong to this specific class of processes. It may, however, be surmised that a number of real phenomena near equilibrium may, with a certain approximation, be adequately described by such Gauss-Markov processes[7]. In section 2.2 we will see that the same results can be achieved for general processes from relations between fluctuation and dissipation in systems far from equilibrium, which have been derived the last years.

To be able to use the results of some "model" stochastic processes given in literature (Ornstein-Uhlenbeck process or damped harmonic oscillator) for a model of thermal or attenuated acoustic waves we will change the initial conditions: instead of a defined initial value (with zero variance) we take the stochastic process at equilibrium at a time zero and before. At the time zero ($t=0$) a certain perturbation is applied to the process, e. g. a rapid change in momentum for the damped harmonic oscillator – called kicked damped oscillator. Reconstruction of the size of this perturbation at time $t=0$ from the measurement after a time $t$ shows how the information about the size of this kick at $t=0$ gets lost with increasing time if diffusive or dissipative processes occur. These new initial conditions have a significant advantage compared to initial conditions used in literature: it turns out that the variance stays constant in time, while the mean value is a function of time. This facilitates the calculations of the entropy production and of the information loss caused by the stochastic process.

The advantage of specifying more precisely the nature of the processes considered is that it enables us to discuss, on the level of the theory of random processes, the behavior of entropy production and of information loss. Following the theory of random fluctuations given e.g. by Groot and Mazur[7], we take as a starting point equations analogous to the Langevin equation used to describe the velocity of a Brownian particle

$$(1) \qquad \frac{d\mathbf{x}}{dt} = -\mathbf{M} \cdot \mathbf{x} + \boldsymbol{\varepsilon}(t).$$

The components of the vector $\mathbf{x}$ are the random variables $x_i$ ($i = 1,2,\ldots,n$), which have at equilibrium a zero mean value. The matrix $\mathbf{M}$ of real phenomenological coefficients is independent of time $t$. The vector $\boldsymbol{\varepsilon}(t)$ represents white noise, which is at different times uncorrelated. The distribution density of $\mathbf{x}$ turns out to be an $n$-dimensional Gaussian distribution

$$(2) \qquad f(\mathbf{x},t) = \frac{1}{\sqrt{(2\pi)^n |\Sigma|}} e^{-\frac{1}{2}(\mathbf{x}-\bar{\mathbf{x}}(t))^T \Sigma^{-1}(\mathbf{x}-\bar{\mathbf{x}}(t))},$$

with the mean value $\bar{\mathbf{x}}(t)$ and the covariance matrix $\Sigma$, which is usually also a function of time $t$. Using equilibrium ($\bar{\mathbf{x}}_0 = 0$) as initial conditions has a significant advantage compared to initial conditions used in literature: it turns out that the covariance stays constant in time, while the mean value is a function of time. This facilitates the calculations of the entropy production and of the information loss due to the stochastic process. $|\Sigma|$ is the determinant of the covariance matrix.

If the initial mean value of $\mathbf{x}$ at time zero is given by $\bar{\mathbf{x}}_0$, the mean value at a later time will be

$$(3) \qquad \bar{\mathbf{x}}(t) = e^{-\mathbf{M}t} \cdot \bar{\mathbf{x}}_0.$$

By an adequate coordinate transformation of $\mathbf{x}$ the matrix $\mathbf{M}$ can be diagonalized: the eigenvalues of $\mathbf{M}$ are the elements of the diagonal matrix.

According to the second law of thermodynamics the entropy of an adiabatically insulated system must increase monotonously until thermodynamic equilibrium is established within the system, where the entropy is set to zero. Then the entropy at a time t is

$$(4) \qquad S(t) = -k_B \int f(\mathbf{x},t) \ln \frac{f(\mathbf{x},t)}{f(\mathbf{x},t \to \infty)} d\mathbf{x} \equiv -k_B D(f(\mathbf{x},t) \| f(\mathbf{x},t \to \infty)),$$

with the Boltzmann constant $k_B$, ln is the logarithm to the base *e*, and *D* is the relative entropy or Kullback-Leibler distance between the distribution at time *t* and the distribution at equilibrium (*t* at infinity). Stein's lemma[25] gives a precise meaning to the relative entropy *D(f||g)* between two distributions: if *n* data from g are given, the probability of guessing incorrectly that the data come from f is bounded by $e^{-nD(f\|g)}$, for *n* large[26]. Therefore, the relative entropy is a quantitative measure of the information loss due to the stochastic process and is equal to the thermodynamic entropy *S(t)*.

For Gaussian distributions with a constant covariance matrix the Kullback-Leibler *D* distance results in (e. g. Horowitz and Jarzynski[27])

$$(5) \qquad D(f(\mathbf{x},t) \| f(\mathbf{x},t \to \infty)) = \frac{1}{2} \overline{\mathbf{x}}(t)^T \Sigma^{-1} \overline{\mathbf{x}}(t).$$

Before modeling heat diffusion and attenuated acoustic waves as a Gauss-Markov process we give simple examples for such processes: the Wiener process to model pure diffusion, the Ornstein-Uhlenbeck process with only one component of **x** as a model for the velocity of a Brownian particle and the damped harmonic oscillator with two components of **x**.

### 2.1.1. Wiener process and random walk

If **M** is zero in eq. (1) the right side of eq. (1) is only white noise and we get the Wiener process, which can be also seen as the limit of a random walk for continuous space (e.g. Gardiner[28]). In one dimension the traditional way to define random walk is to allow the walker to take steps at each time interval at which he must step either backward or forward, with equal probability (Fig. 2). The Wiener process has no stationary state and the distribution density of x is a Gaussian distribution with the mean value constant in time and equal to the starting position, e.g. $\overline{\mathbf{x}} = x(t=0) = 0$ and a variance proportional to the time t. Therefore an initially sharp distribution spreads with increasing time (Fig. 3).

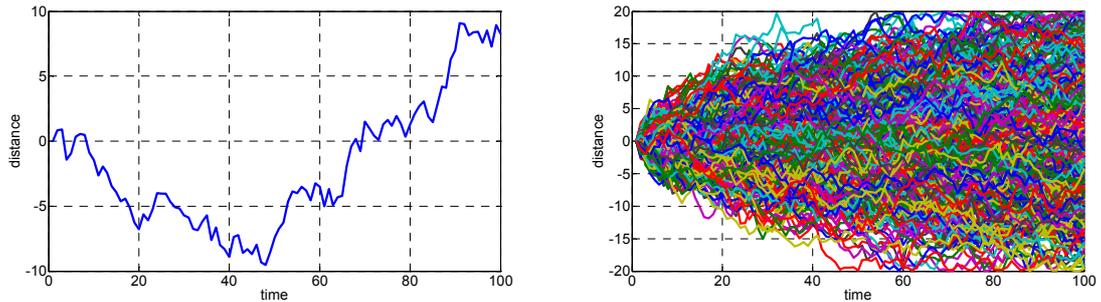

*Fig. 2. One realization of the Wiener process with the particle starting at x=0 (left) and 500 realizations using reflecting boundaries at a distance of ±20 (right).*

For modeling of thermal diffusion not only one walking particle is used but a lot of them. For an efficient modeling instead of the one-particle picture the occupation number representation is used (e. g. Honerkamp[19]). Diffusion is described in this representation as a certain change of occupation numbers $N_i$ of individual cells, which are small volume elements of the sample. In total there are *N* particles, numbered from *j=1, 2, ...N* and therefore $N = \sum_i N_i$.

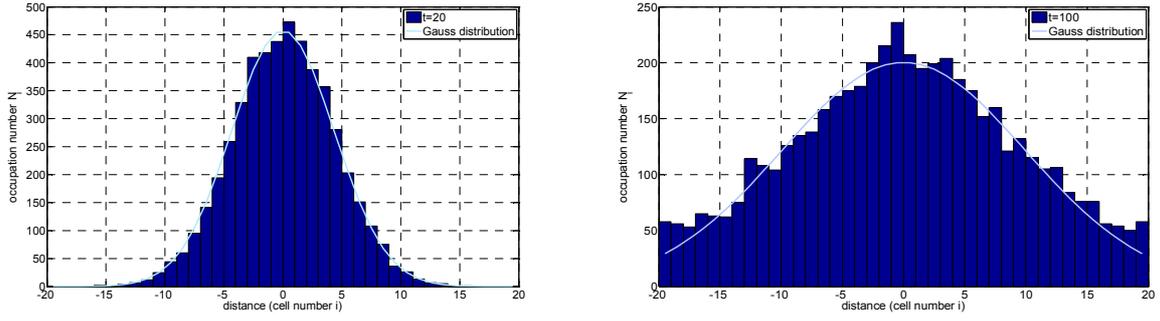

*Fig. 3. Occupation number representation of the Wiener process for 5000 particles starting at x=0 at t=20 (left) and t=100 (right). For comparison the Gaussian distribution with a variance proportional to time is shown. For t=100 the reflecting boundaries at a distance of ±20 can be already recognized by occupation numbers significantly increasing the Gaussian distribution.*

We assume that the diffusion of the particles is independent from each other with $p_i(t)$ giving the probability that a certain particle is in cell $i$ at time $t$ when starting from cell $i=0$ at $t=0$. $p_i(t)$ is the same for every particle and therefore does not depend on $j$. $X_{ij}(t)$ is a new random variable, defined to be one if particle $j$ is in cell $i$ and zero in all other cases. For every particle $X_{ij}(t)$ is equal to one with probability $p_i(t)$, independent from particle number $j$. Therefore the mean value and variance of this random variable are the same for all the particles:

(6)
$$\overline{X}_{ij}(t) = \sum_i X_{ij}(t) p_i(t) = p_i(t); \quad Var(X_{ij}(t)) = \overline{X}_{ij}^2 - (\overline{X}_{ij})^2 = \sum_i X_{ij}^2 p_i - p_i^2 = p_i(t) - p_i(t)^2$$

$$, as\ X_{ij}^2 = X_{ij} = \begin{cases} 1\ if\ particle\ j\ is\ in\ cell\ i \\ 0\ else \end{cases}.$$

With the occupation numbers $N_i = \sum_{j=1}^{N} X_{ij}$ the mean value and variance follows from independence of the particles:

(7) $\quad \overline{N}_i(t) = N p_i(t); \quad Var(N_i(t)) = N p_i(t) - N p_i(t)^2 = \overline{N}_i(t)(1 - p_i(t)) \approx \overline{N}_i(t)\ for\ p_i(t) << 1.$

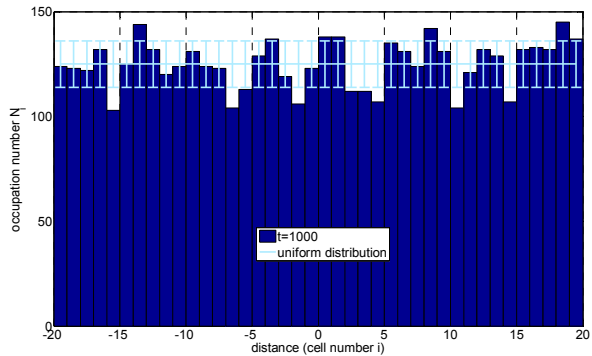

*Fig. 4. Occupation number representation of the Wiener process for N=5000 particles starting at x=0 at t=1000 using 40 cells. At that time the influence of the reflecting boundaries at a distance of ±20 dominates. For comparison the uniform distribution with the mean value 125 (5000 particles divided by 40 cells) and the errorbar at ± the standard deviation is shown. The standard deviation is the square root of the variance given in eq. (7) using $p_i=1/40$ for the uniform distribution.*

In section 3 thermal diffusion is a stochastic process. A simple model for one dimensional heat diffusion using adiabatic boundary conditions is a one dimensional random walk with reflecting boundaries. Such adiabatic boundary conditions can be realized for a thermally isolated sample and the normal derivative of the temperature vanishes at all times at the sample boundaries (see section 3 for more details). The walking particles could be thought as the phonons in the sample volume. Giving them all the same energy is a rough approximation, but the essential details of the influence of fluctuations and the connection of entropy production and information loss can be seen already for such a model system.

Starting from an equilibrium modeled by a uniform distribution (Fig. 4) with $N_{equi}$ particles in $N_{cell}$ cells, at the time zero $N_0$ particles are added at cell $i=0$. Compared to eq. (7) we have now two different distributions: the equilibrium distribution $p^{equi} = 1/N_{cell}$, which is constant in time and space (for every cell $i$), and the Gaussian distribution $p^{Gauss}_i(t)$ for the added particles (at least up to a time when the reflecting boundaries can be still neglected). As the movement of the individual particles is independent one gets for the mean value and variance:

$$(8) \qquad \overline{N}_i(t) = \frac{N_{equi}}{N_{cell}} + N_0 p^{Gauss}_i(t); \quad Var(N_i(t)) = \overline{N}_i(t) - \frac{N_{equi}}{N_{cell}^2} - N_0 p^{Gauss}_i(t)^2 \approx \overline{N}_i(t)$$

An example for $N_{equi}$ =10 000, $N_{cell}$ =40 and $N_0$=1000 is shown in Fig. 5.

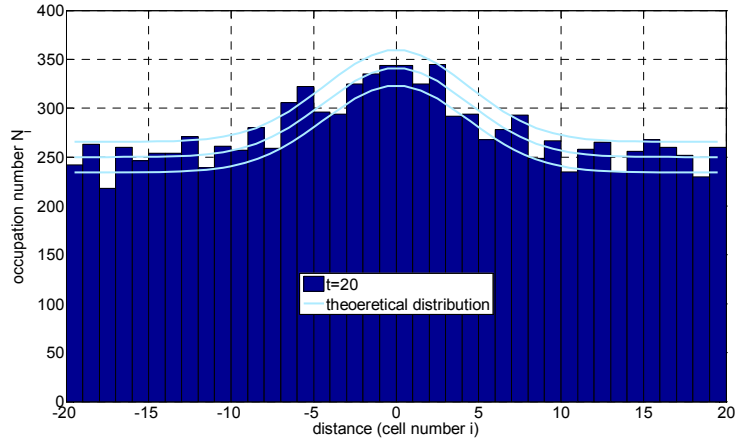

Fig. 5. Occupation number representation of the Wiener process for for $N_{equi}$ =10 000, $N_{cell}$ =40 and $N_0$=1000. For comparison the mean value ± the standard deviation (square root of variance) from eq.(8) is shown.

Despite the independence of the movement of the individual particles the occupation numbers are not uncorrelated. One example is that if a cell has a high occupation number the occupation number of the neighboring cells will increase shortly afterwards. The covariance of the occupation number in cell $i$ at time $t$ and cell $j$ at a later time $t+\tau$ is:

$$(9) \qquad Cov(N_i(t), N_j(t+\tau)) = \sum_{k=1}^{N} Cov(X_{ik}(t), X_{jk}(t+\tau)) = N p_i(t)(p_{i-j}(\tau) - p_j(t+\tau))$$

For the last equation it was used that the probability for a particle to be in cell $j$ at time $t+\tau$ when it was in cell $i$ at time $t$ is $p_{i-j}(\tau)$.

The time development of the occupation numbers $N_i$ is a Gauss-Markov process and can be described by a Master equation (e.g. van Kampen[29]) or by eq. (1). The matrix **M** in eq. (1) is a square matrix with the dimension equal to the number of cells:

$$(10) \quad \mathbf{M} = -\frac{1}{2} \begin{pmatrix} -1 & 1 & 0 & \cdots & 0 & 0 \\ 1 & -2 & 1 & \ddots & & 0 \\ 0 & 1 & -2 & 1 & \ddots & \vdots \\ \vdots & \ddots & \ddots & \ddots & \ddots & 0 \\ 0 & & \ddots & 1 & -2 & 1 \\ 0 & 0 & \cdots & 0 & 1 & -1 \end{pmatrix}$$

With this matrix **M** eq. (1) can be written in components as:

$$(11) \quad dN_i = \frac{1}{2}(N_{i-1} + N_{i+1} - 2N_i)dt + \sqrt{Var\,dt}\,\varepsilon_t(0,1)$$

where $\varepsilon_t(0,1)$ is a random variable with standard distribution (mean value zero and variance 1). The reflecting boundaries cause a change of the first and last line of **M**:

$$(12) \quad \begin{aligned} dN_1 &= \frac{1}{2}(N_2 - N_1)dt + \sqrt{Var\,dt}\,\varepsilon_t(0,1) \\ dN_{N_{cell}} &= \frac{1}{2}(N_{N_{cell}-1} - N_{N_{cell}})dt + \sqrt{Var\,dt}\,\varepsilon_t(0,1) \end{aligned}$$

Eq. (11) can be seen also as a discretized version (finite differences) of the diffusion equation with a diffusion coefficient of $\alpha = 1/2$ (compare eq.(34)).

As mentioned already in eq.(3) the matrix **M** can be diagonalized by an adequate coordinate transform, which turns out to be the Fourier transform in space ("k-space" in section 3). For the discretized random walk with reflecting boundaries this transformation is in fact the discrete cosine transform and the eigenvalues in the diagonal matrix in k-space are:

$$(13) \quad \gamma_k = 4\sin^2(k\pi/2) \approx \pi^2 k^2 \text{ with } k = 0,1,\ldots,N_{cell}-1.$$

The last approximation is for a continuous k-space when the cell volume gets zero and $N_{cell}$ gets to infinity. These eigenvalues are proportional to the square of the wavenumber k and are the time constants for the decay in time. Therefore variations with a high wavenumber (low wavelength) decay fast. Wavenumber $k=0$ shows no decay and represents the equilibrium temperature, which will be reached after a long time. The stochastic process in k-space with the diagonalized matrix can be split up into several independent stochastic processes with one-dimensional random variables. Each such process is an Ornstein-Uhlenbeck process, which will be described in detail in the next section.

### 2.1.2. Kicked Ornstein-Uhlenbeck process

If the random vector **x** in eq. (1) has only one component we get the Langevin equation

$$(14) \quad \frac{dv(t)}{dt} = -\gamma \cdot v(t) + \sigma\,\eta(t)$$

which was used to describe Brownian motion of a particle. The random variable $v$ is the particle velocity, $-\gamma \cdot v$ is the viscous drag, and $\sigma$ is the amplitude of the random fluctuations. The Langevin equation governs an Ornstein-Uhlenbeck process, after L. S. Ornstein and G. E. Uhlenbeck, who formalized the properties of this continuous Markov process[30]. We assume that initially we have thermal equilibrium with zero mean velocity and at time zero the particle is kicked, which causes an immediate change in velocity of $v_0$. Following eq. (3) the mean value $\bar{v}(t)$ shows an exponential decay:

(15) $$\bar{v}(t) = e^{-\gamma t} \cdot \bar{v}_0$$

The variance of the velocity turns out to stay constant in time and is therefore equal to the equilibrium value $Var(v) = \sigma^2/2\gamma$ (e.g. Honerkamp[19]). In Fig. 6 time and velocity are scaled to be dimensionless and the standard deviation (square root of the variance) of the velocity is normalized.

From eq. (5) the information loss equal to the entropy production till the time t after the kick is:

(16) $$\Delta S(t) = k_B \frac{\gamma}{\sigma^2} \bar{v}(t)^2$$

On the other hand the entropy production known from thermodynamics is the dissipated energy $\Delta Q$, which is the kinetic energy of the Brownian particle of mass m, divided by the temperature T:

(17) $$\Delta S(t) = \frac{\Delta Q}{T} = \frac{m \bar{v}(t)^2}{2T}$$

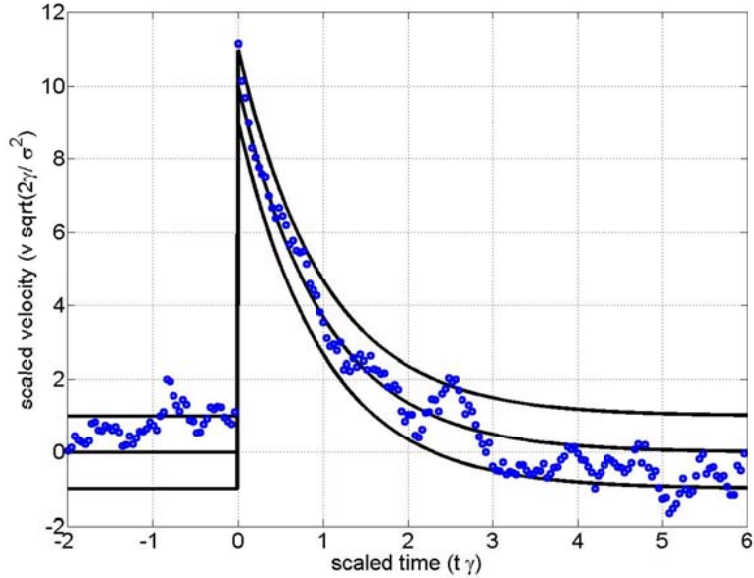

*Fig. 6. Points on a sample path of the normalized kicked Ornstein-Uhlenbeck process defined by the Langevin eq. (14). The solid lines represent the mean, and mean ± standard deviation of the scaled velocity coordinate. At the time t=0 a value of $v_0$=10 has been added to the scaled velocity. After some time the information of the amplitude gets more and more lost due to the fluctuations.*

The thermodynamic entropy production in eq. (17) has to be equal to the loss of information in eq. (16), and therefore we get for the variance of the velocity:

(18) $$\frac{\sigma^2}{2\gamma} = \frac{k_B T}{m}$$

For equilibrium eq. (18) has been derived in the past by using the equipartition theorem, which states that the equilibrium energy associated with fluctuations in each degree of freedom is $k_B T/2$. We have used the equity of entropy production and information loss. Eq. (18) states a connection between the strength of the fluctuations, given by $\sigma^2$, and the strength of the dissipation $\gamma$. This is the fluctuation-dissipation theorem (FDT) in its simplest form for uncorrelated white noise and this derivation of eq. (18) shows the information theoretical background of the FDT.

It is instructive to determine the least square estimator[19] for the kicking velocity $v_0$. If we write for the estimated (reconstructed) kicking velocity $v_r$

$$(19) \qquad v_r = R(t) \cdot v(t)$$

we calculate $R(t)$ by minimizing the mean error $\langle (v_0 - v_r)^2 \rangle$:

$$(20) \qquad R(t) = \frac{e^{-\gamma t}}{e^{-2\gamma t} + \sigma^2 / 2\gamma v_0^2}$$

This gives the Tikhonov regularization[31] with $\sigma^2 / 2\gamma v_0^2 = Var(v)/v_0^2$ as regularization parameter (see also section 3.3 for more information on regularization methods). The inverse square root of the regularization parameter $v_0 / \sqrt{Var(v)}$ can be interpreted as signal-to-noise-ratio (SNR) of $v(t)$. This least square estimator for the kicking velocity $v_0$ is a biased estimator as the expectation value is $R(t)e^{-\gamma t}v_0 < v_0$.

Another example for an Ornstein-Uhlenbeck process using diffusion instead of dissipation is the distribution of *2N* particles between two boxes of volume *V*, which are connected by a small hole where particles can slip through between the two boxes (effusion). The constant time rate $\gamma$ is the reciprocal value of the mean time of a particle to stay in one of the boxes before it slips through the hole. At a certain time *t*, $N + \delta N(t)$ particles are situated in the left box and $N - \delta N(t)$ particles are in the right box (Fig. 7).

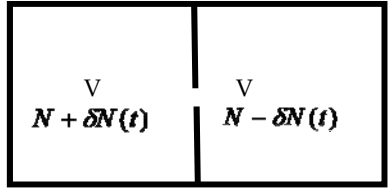

*Fig. 7. Distribution of 2N particles between two boxes of volume V, which are connected by a small hole where particles can change between the two boxes at a constant time rate $\gamma$ (effusion). At a certain time t, $N + \delta N(t)$ particles are situated in the left box and $N - \delta N(t)$ particles are in the right box.*

$\delta N(t)$ can be approximated by an Ornstein-Uhlenbeck process for a big number of particles (see e.g. Lemons[20]). The mean value $\overline{\delta N}(t)$ is given by an exponential decay in time according to eq. (15). For the variance $\sigma^2 / 2\gamma$ the equipartition theorem cannot be used, because the effusion process is not dissipative in the sense that energy is converted into heat. The information loss is still described analog to eq. (16) by:

$$(21) \qquad \Delta S(t) = k_B \frac{\gamma}{\sigma^2} \overline{\delta N}(t)^2$$

Spatial diffusion is the cause for the entropy production and is described by $S = k_B \ln W$. *W* is the number of possibilities to distribute the particles into the two boxes for a certain $\overline{\delta N}$. Using Stirlings formula for a big number of particles one gets for the entropy production:

(22) $$\Delta S(t) = k_B \frac{1}{N} \overline{\delta N}(t)^2$$

By comparison with eq. (21) one gets again a relation between the fluctuations, given by $\sigma^2$, and the strength of the diffusion, given by the rate $\gamma$ (fluctuation-"diffusion" theorem) and for the variance of $\delta N$ we get:

(23) $$Var(\delta N) = \frac{\sigma^2}{2\gamma} = \frac{N}{2}$$

### 2.1.3. Kicked harmonic oscillator

For modeling of acoustic waves one needs in addition to dissipation an oscillating term. For pure oscillation without damping we have no loss of information (Fig. 8).

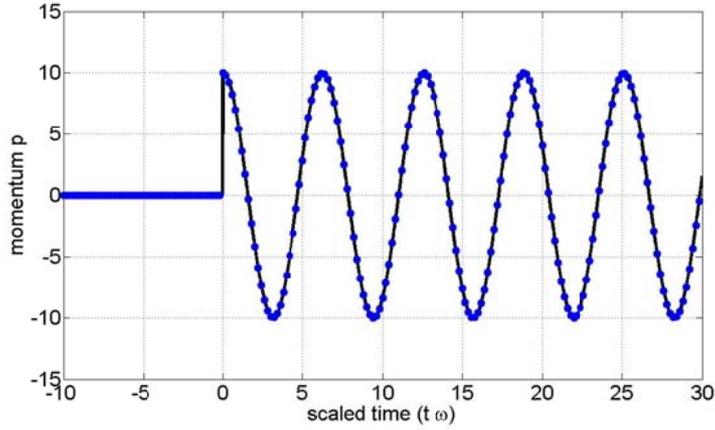

*Fig. 8. Points on a sample path of the kicked harmonic oscillator without damping. The momentum used to kick the oscillator can be reconstructed without loss of information at any time t after the kick if ω is known.*

The stochastically damped harmonic oscillator combines the oscillatory and the diffusive behavior and therefore is a good starting point to model attenuated acoustic waves. The equations of motion are (using the momentum $p$ instead of the velocity $v$):

(24) $$\frac{dx(t)}{dt} = \frac{1}{m} p(t)$$

(25) $$\frac{dp(t)}{dt} = -\frac{\gamma}{m} \cdot p(t) - m\omega_0^2 x + \sigma\, \eta(t)$$

These equations can be combined using a two dimensional random vector $\mathbf{x} = (x, p)$ (eq. (1)) and were solved already 1943 by Chandrasekhar[32] for definite initial conditions $x(0)$ and $p(0)$. Again we have changed the initial conditions to an oscillator with zero mean values kicked by an initial momentum $p_0$ at time zero. In Fig. 9 the damping is chosen $\gamma = m\omega_0/3$.

Using the fluctuation-dissipation theorem $\sigma^2/2\gamma = k_B T$ one obtains for the distribution function (eq. (2))

(26) $$f(x,p,t) = \frac{1}{2\pi \frac{k_B T}{\omega_0}} \exp(-\frac{1}{2mk_B T}(p - \overline{p}(t))^2 - \frac{1}{2k_B T} m\omega_0^2 (x - \overline{x}(t))^2)\ ,$$

where $\bar{x}(t)$ and $\bar{p}(t)$ are the solutions of the ordinary (non-stochastic) damped harmonic oscillator. From eq. (5) then one gets for the information loss

$$(27) \qquad S(t) = -\frac{1}{T}(\frac{1}{2}m\omega_0^2 \bar{x}(t)^2 + \frac{\bar{p}(t)^2}{2m}) = -\frac{1}{T}(E_{pot} + E_{kin}),$$

which is equal to the entropy from thermodynamics, where $E_{pot}+E_{kin}$ is the total energy of the harmonic oscillator (sum of the potential and kinetic energy). This fact confirms again that the fluctuation-dissipation theorem for the damped harmonic oscillator $\sigma^2/2\gamma = k_B T$ can be derived from the equity of entropy production and information loss.

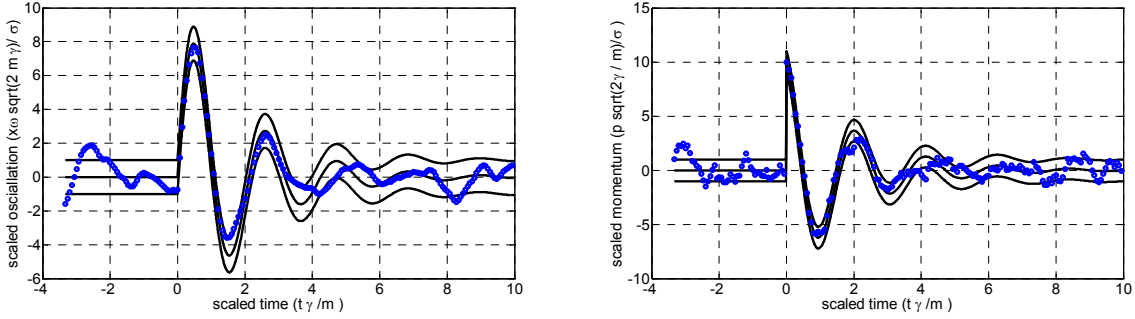

*Fig. 9. Points on a sample path of the normalized kicked damped harmonic oscillator (eq. (24) and (25)). The solid lines represent the mean, and mean ± standard deviation of the scaled oscillation (left) and momentum (right). At the time t=0 a value of $p_0$=10 has been added to the scaled momentum. After some time the information about the value of $p_0$ gets more and more lost due to the fluctuations.*

For the mean value of the momentum $\bar{p}(t)$ (and also $\bar{x}(t)$) one gets from eq. (3):

$$(28) \qquad \bar{p}(t) = p_0[\cos(\omega t) - \frac{\gamma}{2m}\frac{\sin(\omega t)}{\omega}]e^{-\frac{\gamma t}{2m}}$$

Compared to the Ornstein-Uhlenbeck process (eq. (15)) the mean value has not only an exponential decay in time but also shows oscillations with a frequency $\omega = \sqrt{\omega_0^2 - \gamma^2/(4m^2)}$. As mentioned above (Fig. 8) the oscillating term does not change the entropy and no information is lost. In the average only the exponential decay causes production of entropy and the information about the value of $p_0$ gets more and more lost due to the fluctuations. The entropy production which is proportional to the total energy is shown in *Fig. 10*.

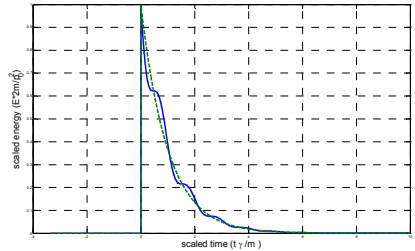

*Fig. 10. Total energy of the damped harmonic oscillator (solid line) shows in the average an exponential decay with the time constant $\gamma / m$ (dashed line).*

The damped harmonic oscillator can be reduced to a simple one dimensional Ornstein-Uhlenbeck process in the overdamped limit, in which the momentum effectively equilibrates instantaneously, and as a result the momentum does not contribute to the entropy. By setting eq. (25) to zero and putting this into eq. (24) one gets

(29)
$$\gamma \frac{dx(t)}{dt} = -m\omega_0^2 x + \sigma \eta(t),$$

which gives in an exponential decay with the time constant $m\omega_0^2/\gamma$ and variance $\frac{k_B T}{m\omega_0^2}$ for $\bar{x}(t)$.

## 2.2. Entropy production for general stochastic processes

In section 2.1 the stochastic process was assumed to have Gauss-Markov character. Now we will show that the same results can be achieved for general processes from relations between fluctuation and dissipation in systems far from equilibrium, which have been derived the past years[26,33,34]: Jarzynski described a "forward" process starting from an equilibrium state at a temperature $T$, during which a system evolves in time as a control parameter $\lambda$ is varied from an initial value $A$ to a final value $B$. $W$ is the external work performed on the system during one realization of the process; $\Delta F = F_B - F_A$ is the free energy difference between two equilibrium states of the system, corresponding to $\lambda = A$ and $B$. The "reverse" process starts from an equilibrium state with $\lambda = B$ and evolves to $\lambda = A$. For the relative entropy $D$ between the forward and reverse process it was shown that

(30)
$$\Delta S = \frac{\langle W \rangle - \Delta F}{T} \geq k_B D(f_A \| f_B),$$

with the average performed work $\langle W \rangle$. The probability density in phase space for the equilibrium state with $\lambda = A$ (forward process) is called $f_A$ and $f_B$ is the phase space density for the for the equilibrium state with $\lambda = B$ (reverse process).

For an instantaneous change of the control parameter $\lambda$ the equality sign holds[34] in eq. (30) and this gives the same result as eq. (4) in section 2.1, but not only for Gauss-Markov processes but for general stochastic processes. Using the Boltzmann probability distributions for the equilibrium distributions $f_A \equiv 1/Z_A \exp(-H(x;A)/k_B T)$ and $f_B \equiv 1/Z_B \exp(-H(x;B)/k_B T)$ this can be directly derived (details can be found in Kawai et al.[34]). $H(x;\lambda)$ is the Hamiltonian, where $x=(q,p)$ represents the set of position and momentum variables. $Z_\lambda$ is the normalization factor (partition function) and $\Delta F = -k_B T \ln(Z_B/Z_A)$. For the averaged performed work $\langle W \rangle$ one gets

(31)
$$\langle W \rangle = \int dx [H(x;B) - H(x;A)] f(x),$$

since the state $x$ and therefore the distribution $f(x)$ does not change during the instantaneous change of the control parameter. From eq. (30) the entropy production for the non-equilibrium process $\Delta S$ is the average dissipated work $\langle W \rangle - \Delta F$ divided by the equilibrium temperature $T$. This entropy production is equal to the information loss for an instantaneous change of the control parameter. *The information loss is the right side of eq. (30) and the equality to entropy production on the left side is essential for an information theoretical treatment of non-destructive imaging.*

In the rest of this section we show that particularly for Gaussian distributions the general eq. (30) will give the same results as for the linearized theory in section 2.1. For the linearized theory the stochastic process started also from equilibrium and was "kicked" at $t = 0$. For example for a Brownian particle described as kicked Ornstein-Uhlenbeck process in section 2.1.2, the velocity of the particle was suddenly increased by $v_0$. The same state just after the kick can be reached as an equilibrium state with a constant external force $F = m \cdot \gamma \cdot v_0$ applied to the particle and from eq (14) one gets

(32)
$$\frac{dv(t)}{dt} = -\gamma \cdot v(t) + \gamma \cdot v_0 + \sigma \, \eta(t).$$

The external force $F$ is the control parameter $\lambda$ from above. For the forward process this force is suddenly reduced to zero at $t = 0$. In the reverse process this force is switched on. Therefore the right side of eq. (30) is equal to the right side of eq. (4) (only the sign is changed for the entropy production $\Delta S$ compared to the entropy $S$ in eq. (4)):

(33)
$$k_B D(f(\bar{v} = v_0) \| f(\bar{v} = 0)).$$

The left side of eq. (30) contains the average performed work $\langle W \rangle$ and the free energy difference $\Delta F$. For the average performed work we get from eq. (31) the additional kinetic energy generated by the kick of the Brownian particle $m\bar{v}_0^2/2$. The free energy does not change after the kick. Therefore eq. (17) and eq. (18) follows directly from eq. (30) using the equality sign for an instantaneous change of the control parameter $\lambda$.

In summary in section 2 the equity of thermodynamic entropy production and information loss – measured as the relative entropy $D$ - was shown for general non-equilibrium processes. The underlying assumptions are that we start with an equilibrium process and the "disturbance" at time $t=0$ has to be short, but it has not to be small. For thermography or photoacoustic imaging these conditions are fulfilled if the initial pulse is short compared to the time needed for the diffusion of heat or the propagation of the acoustic wave along a distance in the range of the size of the imaged structures, called thermal or stress confinement[35], respectively. Therefore the equity of thermodynamic entropy production and information loss will be applied to thermal diffusion after such a short pulse in the next section.

## 3. Thermal diffusion as a stochastic process
### 3.1. Thermal diffusion equation in real and Fourier space

Thermal diffusion can be described by the differential equation (Fourier, 1823, or e.g. Rosencwaig[2])

$$(34) \quad (\Delta - \frac{1}{\alpha}\frac{\partial}{\partial t})T(\mathbf{r},t) = -\frac{1}{\kappa}q(\mathbf{r},t),$$

where $T$ is the temperature and $q$ is the thermal source volumetric density as a function of space $\mathbf{r}$ and time $t$. $\Delta$ is the Laplacian operator. $\alpha$ is the thermal diffusivity and $\kappa$ is the thermal conductivity, which are both material properties.

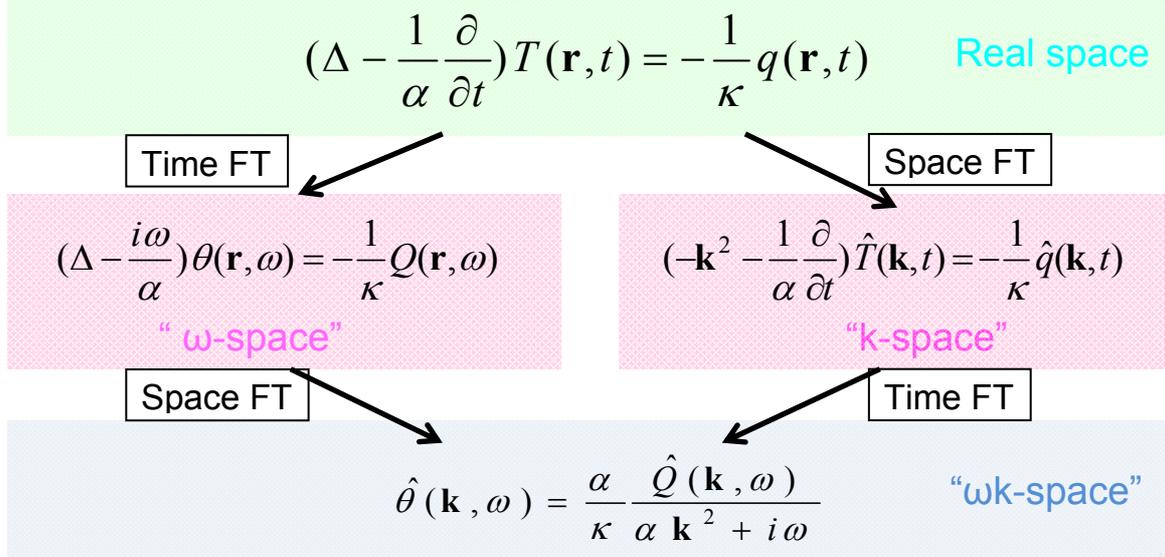

*Fig. 11. The diffusion equation (34) can be solved by Fourier transform (FT) in time (eq. (35)) and a subsequent FT in space (eq. (37)). The same result in "ωk-space" is gained if the space FT is performed first, and then the time FT.*

In the very well developed theory of heat transfer (e. g. by Carslaw and Jaeger[36]) usually a bilateral Fourier transform is applied to $T$ and $q$ in the diffusion equation

$$(35) \quad \begin{aligned} g(t) &= \frac{1}{2\pi}\int_{-\infty}^{+\infty} G(\omega)\exp(i\omega t)\,d\omega \\ G(\omega) &= \int_{-\infty}^{+\infty} g(t)\exp(-i\omega t)\,dt \end{aligned},$$

where $i=\sqrt{-1}$ and ω is angular frequency. When eq. (35) is applied to $T$ and $q$ in eq. (34), one gets for the Fourier transformed $\theta$ and $Q$ the inhomogeneous form of the Helmholtz equation

$$(36) \quad (\Delta - \frac{i\omega}{\alpha})\theta(\mathbf{r},\omega) = -\frac{1}{\kappa}Q(\mathbf{r},\omega).$$

This equation may be solved by the application of a bilateral Fourier transform over space, defined as

(37)
$$G(\mathbf{r}) = \frac{1}{(2\pi)^3} \iiint \hat{G}(\mathbf{k}) \exp(-i\mathbf{kr}) d\mathbf{k},$$
$$\hat{G}(\mathbf{k}) = \iiint G(\mathbf{r}) \exp(i\mathbf{kr}) d\mathbf{r}$$

where $\mathbf{k}$ is the wave vector (see e.g. M. Buckingham[37]). When eq. (37) is applied to $\theta$ and $Q$ of eq. (36), the Helmholtz equation reduces to

(38)
$$\hat{\theta}(\mathbf{k},\omega) = \frac{\alpha}{\kappa} \frac{\hat{Q}(\mathbf{k},\omega)}{\alpha \mathbf{k}^2 + i\omega}.$$

The same result can be achieved if first the spatial Fourier transform (37) is applied to $T$ and $q$ in eq. (34),

(39)
$$(-\mathbf{k}^2 - \frac{1}{\alpha}\frac{\partial}{\partial t})\hat{T}(\mathbf{k},t) = -\frac{1}{\kappa}\hat{q}(\mathbf{k},t),$$

and then the temporal Fourier transform (35) is applied to eq. (39), as shown in Fig. 11.

In the past most publications have used the Helmholtz equation ("ω-space" in the left of Fig. 11). For example Mandelis has solved this equation in his book "Diffusion-Wave Fields"1 in one, two, or three dimensions and for different boundary conditions (e.g. adiabatic boundary conditions where no heat can be "lost" at the sample boundary) by using Green functions. For a delta source in space and time and no boundaries one gets in one dimension

(40)
$$T(x,t) = \frac{1}{4\pi} \int_{-\infty}^{+\infty} \exp(i\omega t) \frac{1}{\sqrt{i\omega\alpha}} \exp(-\sqrt{\frac{i\omega}{\alpha}}|x|) d\omega$$
$$= \frac{1}{\pi} \int_{0}^{+\infty} \cos(\omega t) \frac{1}{\sqrt{\omega\alpha}} \exp(-\sqrt{\frac{\omega}{2\alpha}}|x|) \cos(\sqrt{\frac{\omega}{2\alpha}}|x| + \frac{\pi}{4}) d\omega = \frac{1}{2\sqrt{\pi\alpha t}} \exp(-\frac{x^2}{4\alpha t}).$$

The thermal wave is a superposition of waves with angular frequency $\omega$ and $\sqrt{\omega/2\alpha}$ as wavenumber. At a distance $x$ the amplitude is damped by $\exp(-x\sqrt{\omega/2\alpha})$. Therefore the damping factor after a distance of one wavelength is $\exp(-2\pi) \approx 1/535$. This is the reason why thermal waves are a good example for "very dispersive" waves.

In "k-space" the thermal wave is a superposition of waves damped in time (see e.g. Buckingham[37]). For a delta source in space and time and no boundaries one gets in one dimension:

(41)
$$T(x,t) = \frac{1}{2\pi} \int_{-\infty}^{+\infty} \exp(-ikx) \exp(-k^2 \alpha t) dk = \frac{1}{2\sqrt{\pi\alpha t}} \exp(-\frac{x^2}{4\alpha t}).$$

Of course the result is the same as in eq. (40). These two different representations of a damped wave correspond to a real frequency and a complex wave vector or vice versa (compare to Roitner et al. for attenuated acoustic waves[38]). In the next section the exponential damping in time in eq. (41) will be realized by an Ornstein-Uhlenbeck process, as described in section 2.1.2.

### 3.2. Thermal diffusion in k-space

At a time $t = 0$, an initial temperature distribution $T_0(\mathbf{r})$ is given on the whole sample volume $V$. The sample is thermally isolated, which results in adiabatic boundary conditions: the normal derivative of the temperature $T$ at the sample boundary vanishes at all times $t$. Therefore after a long time $t$ a thermal equilibrium will establish with an equilibrium temperature $\bar{T}$.

For numerical calculations we use a discrete space $T_j = T(\mathbf{r}_j)$, where $\mathbf{r}_j$ are $N$ points on a cubic lattice with a spacing of $\Delta r$ within the sample volume $V$ ($j=1,...,N$). At a time $t$ the temperature distribution $T_j(t)$ can be represented by the Fourier series[39] (including the time $t = 0$ with the known initial temperature distribution $T_0(\mathbf{r})$):

(42) $$T_j(t) = T(\mathbf{r}_j, t) = \sum_{k=0}^{N-1} \hat{T}_k(t) \phi_k(\mathbf{r}_j) \text{ with } \phi_k(\mathbf{r}) = e^{2\pi i \boldsymbol{\rho}_k \cdot \mathbf{r}} \cdot I(\mathbf{r}).$$

$I(\mathbf{r})$ is a support function which is one within the sample volume $V$ and zero outside. $\boldsymbol{\rho}_k$ are integer points on an infinite 3D lattice in k-space. The index $k = 0$ should correspond to $\boldsymbol{\rho}_k = \mathbf{0}$, and therefore $\hat{T}_0(t)$ is equal to the equilibrium temperature $\bar{T}$, which is constant in time. From the diffusion equation (34) we get

(43) $$\hat{T}_k(t) = \hat{T}_k(0) \exp(-\gamma_k t) \text{ with } \gamma_k = 4\pi^2 \alpha \boldsymbol{\rho}_k^2.$$

In Fourier space the time evolution is just an exponential decay in time (Fig. 12). Only for $k = 0$, where $\boldsymbol{\rho}_k = \mathbf{0}$, the Fourier coefficient is constant in time and is equal to the equilibrium temperature as stated above.

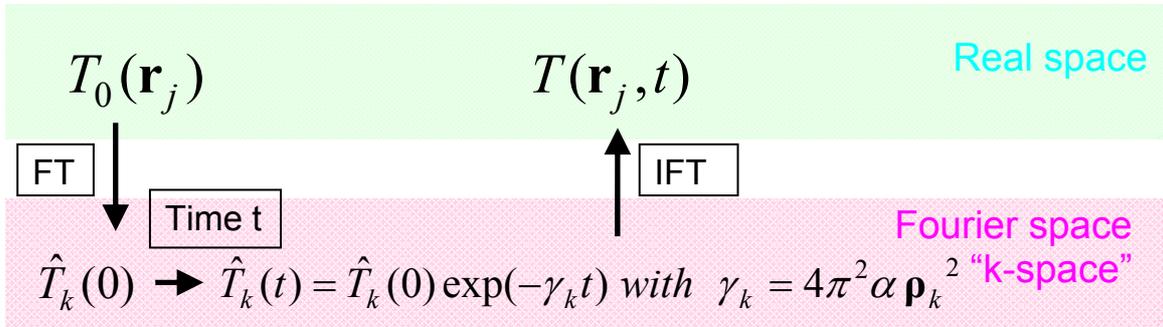

Fig. 12. The initial temperature distribution $T_0(\mathbf{r})$ just after the laser pulse is Fourier transformed (FT). The time evolution of the Fourier series coefficients can be described similar to the mean value of an Ornstein-Uhlenbeck process (eq. (15)). The temperature distribution $T(\mathbf{r},t)$ after a time $t$ is then calculated by an inverse Fourier transform (IFT).

Now we assume that in Fourier space the temporal evaluation of the temperature $\hat{T}_k$ is realized by a Gauss-Markov process. The coordinate transformation, which diagonalizes the matrix $\mathbf{M}$ from eq. (3), is the Fourier transform eq.(42). The elements of the diagonalized matrix $\mathbf{M}$ are the time constants $\gamma_k$ for the exponential decay of $\overline{\hat{T}_k}(t)$, which increase with higher order of $k$ (quadratic with length of $\boldsymbol{\rho}_k$). To get the variance $s_k^2$ of the Gaussian distribution for $T_k(t)$ we use again the fact that the loss of information has to be equal to the thermodynamical entropy production. From eq. (5) we get for the loss of information

(44) $$\Delta S(t) = \frac{1}{2} k_B \sum_{k=1}^{N-1} \frac{1}{s_k^2} \overline{\hat{T}_k}(t)^2.$$

The term with $k = 0$ for the equilibrium temperature is subtracted and therefore the sum starts with $k = 1$ to ensure that the information loss is zero for the equilibrium distribution. Compared to eq. (5) the sign is changed as the entropy production is positive.

On the other hand one gets from thermodynamics for the entropy production of a discrete temperature distribution around the equilibrium value $\bar{T}$ in real space[7] and in Fourier space

$$(45) \quad \Delta S(t) = \frac{C_V}{2\overline{T}^2} \sum_{i=1}^{N} \delta V_i (T_i - \overline{T})^2 = \frac{C_V}{2\overline{T}^2} V \sum_{k=1}^{N-1} \overline{\hat{T}_k(t)}^2,$$

with $\delta V_i$ is the volume of volume element number $i$. Their sum is the sample Volume $V$. $C_V$ is the heat capacity at a constant volume. Comparison with eq. (44) shows that the variance in Fourier space is the same for all wavenumbers with index $k$:

$$(46) \quad s_k^2 = k_B \frac{\overline{T}^2}{C_V V}.$$

For each wave number index $k$, $\gamma_k$ in eq. (43) gives the strength of dissipation and $s_k\sqrt{2\gamma_k}$ gives the amplitude of the random fluctuations for the Ornstein-Uhlenbeck process for $\hat{T}_k(t)$. The stochastic differential equation according to eq. (14) is

$$(47) \quad \frac{d\hat{T}_k(t)}{dt} = -\gamma_k \cdot \hat{T}_k(t) + s_k\sqrt{2\gamma_k}\, \eta(t); \quad k = 1,...,N-1.$$

Fig. 13 and Fig. 14 show a one dimensional example for $N = 400$. The initial temperature distribution is Gaussian shaped with a maximum of 600 Kelvin (K) at a depth of 200 and the full width at 1/e of the maximum is 16 arb. units. The minimum temperature is 300 K. The deviation from equilibrium temperature $\Delta T = T - \overline{T}$ is shown in black in Fig. 13 in real space and in Fig. 14 in Fourier space ("k-space"). The thermal diffusivity is chosen as 50. The temperature profiles at the 400 discrete points are calculated by 400 independent Orstein-Uhlenbeck processes for $\hat{T}_k(t)$ at a time $t=1$, $t=5$, and $t=20$. In Fourier space higher orders of k show a rapid decrease. The Fourier coefficients are scaled to give a variance $s_k^2 = 1$. The standard deviation in real space is $s_k\sqrt{(N-1)/2}$, which gives 14.12 K and is constant in depth and time, as the initial state is the temperature distribution of the equilibrium state increased by $T_0$ (analog to the kicked Ornstein-Uhlenbeck process). The standard deviation is selected rather high for this simulation to show clearly the influence of fluctuations.

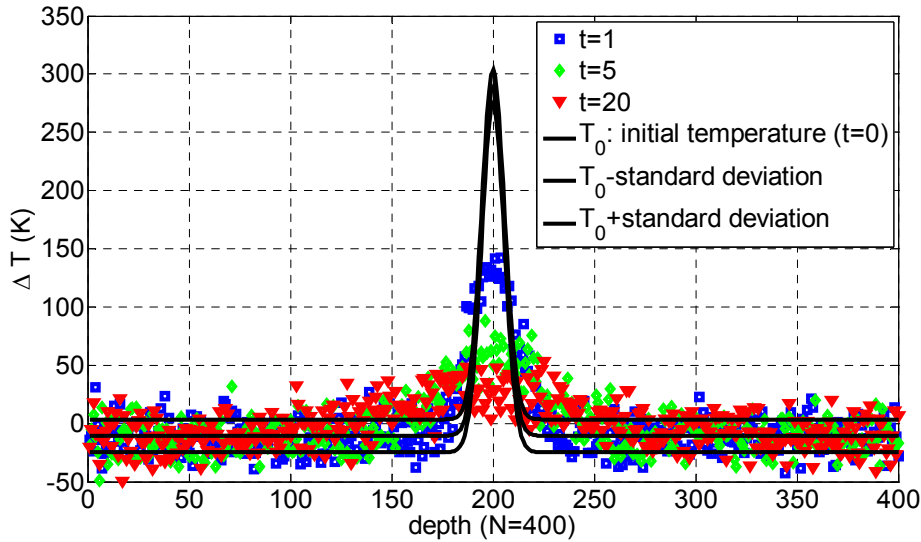

*Fig. 13. Deviation from equilibrium temperature $\Delta T = T - \overline{T}$ in real space as a function of time. The initial temperature distribution is Gaussian shaped with a maximum of 600 Kelvin (K) at a depth of 200 and the full width at 1/e of the maximum is 16 arb. units. The minimum temperature is 300 K. The standard deviation in real space (see text) is constant in depth and time.*

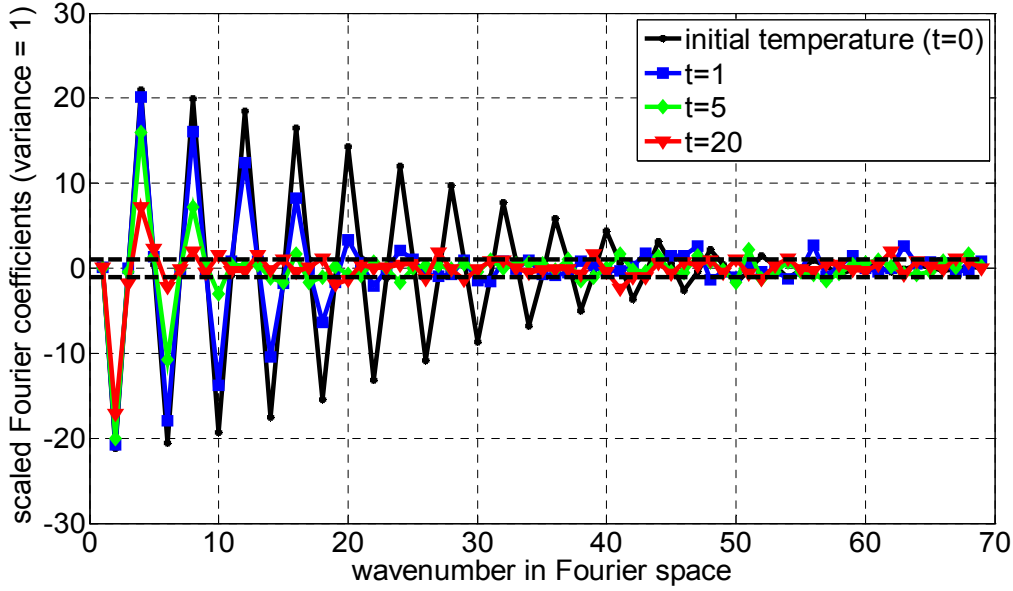

*Fig. 14. Fourier coefficients as a function of time. Coefficients with a higher wavenumber show a rapid decay with time according to eq. (43). The Fourier coefficients are scaled to have a variance of one (independent of wavenumber).*

After a long time the mean value of all Fourier coefficients will vanish. Only the Fourier coefficient for $k=0$ is constant and equal to the equilibrium temperature. The entropy from eq. (44) gets zero and all the information about the shape of the initial temperature profile is lost. For earlier times the temperature profile is broadened (Fig. 13) but the initial temperature can be reconstructed to a certain extent, which depends on the fluctuations $s_k\sqrt{2\gamma_k}$. How to get the best reconstruction of the initial temperature profile from a "measured" temperature distribution at a later time $t$ will be the topic of the next section.

### 3.3. Inverse thermal diffusion and regularization methods in k-space

In the previous section the initial temperature distribution $T_0(\mathbf{r})$ was given and the temperature distribution $T(\mathbf{r},t)$) after a time $t$ was calculated (Fig. 12). Now we want to reconstruct the initial temperature distribution from the measured temperature distribution at a time $t$ (inverse problem). Without fluctuations ($s_k = 0$) this reconstruction in k-space would be (from eq. (43)):

$$(48) \qquad \hat{T}_k(0) = \hat{T}_k(t)\exp(+\gamma_k t) \text{ with } \gamma_k = 4\pi^2\alpha\,\mathbf{\rho}_k^{\,2}.$$

These exponential time evolution factors can reach high values for increasing length of wave vectors $\mathbf{\rho}_k$. Then even small fluctuations are "blown up" exponentially and the reconstruction error is big. Therefore regularization methods have to be used for this inverse problem (see e.g. Hansen[31]), like truncated singular value decomposition (SVD) or Tikhonov regularization[40]. For truncated SVD only the first $i$ wavenumbers are taken and the other wavenumbers for which the exponential factors produce high errors are set to zero:

$$(49) \qquad \hat{T}_k(0) = \begin{cases} \hat{T}_k(t)\exp(+\gamma_k t), & \text{for } k \leq i \\ 0 & \text{otherwise} \end{cases}.$$

Tikhonov regularization looks for a $\hat{\mathbf{T}}(0)$ which fulfills the linear vector equation $\hat{\mathbf{T}}(t) = \hat{\mathbf{T}}(0)\exp(-\gamma t)$ as good as possible, but is also a smooth solution, minimizing $\hat{\mathbf{T}}(0)^2$. The regularization parameter $\lambda$ gives the tradeoff between these two requirements:

$$\min\{((\hat{\mathbf{T}}(t) - \hat{\mathbf{T}}(0)\exp(-\boldsymbol{\gamma}t))^2 + \lambda\hat{\mathbf{T}}(0)^2\}. \quad (50)$$

Differentiating with respect to $\hat{T}_k(0)$ and setting the derivative equal to zero results in

$$\hat{T}_k(0) = \frac{\exp(-\gamma_k t)}{\exp(-2\gamma_k t) + \lambda} \hat{T}_k(t). \quad (51)$$

Compared to the SVD method (eq. (49)) this will give the same $\hat{T}_k(0)$ for $k << i$ or for $k >> i$. In the interim region ($k \sim i$) the Tikhonov regularization does not show a cut at $k = i$ but a smooth transition. Fig. 15 shows this for the example of the temperature profile given in Fig. 13 for a time $t=20$. At that time wave numbers increasing $k=8$ are determined by the fluctuations and have nearly no additional information about the initial temperature profile (Fig. 14).

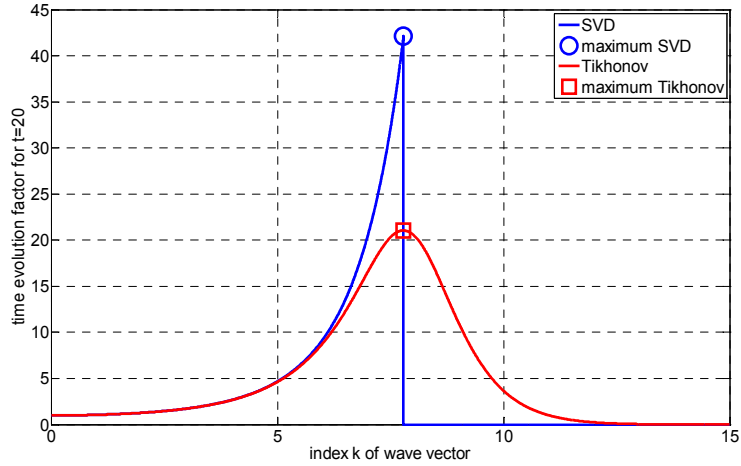

*Fig. 15. Time evolution factor for the initial temperature profile shown in Fig. 13 for a time t=20. At that time wave numbers increasing k=8 are determined by the fluctuations and have nearly no additional information about the initial temperature profile (Fig. 14).*

The Tikhonov time evolution factor at a certain time $t$ has its maximum for the wave vector with index $i$ if

$$\lambda = \exp(-2\gamma_i t). \quad (52)$$

A comparison of the time evolution factor between SVD and Tikhonov reconstruction (eq. (49) and (51) ) is shown in Fig. 15. The signal-to-noise-ratio (SNR) of the measured temperature profile at the time $t$ determines how many wave vectors can be used for the reconstruction of the initial temperature profile. Compare with eq. (20), where the Tikhonov regularization is the least square estimator with the regularization parameter $\lambda$ as the variance divided by signal amplitude, which is the square of the reciprocal value of the SNR:

$$\exp(\gamma_i t) = SNR. \quad (53)$$

If the SNR gets higher (which means less noise), then more Fourier coefficients can be reconstructed (higher $i$) at a certain time $t$ and the reconstructed profiles will contain more information (e.g. less FWHM for Gaussian profiles). The reconstructions using the time evolution factors from Fig. 15 for SVD and Tikhonov reconstruction are shown in Fig. 16. Tikhonov reconstruction is slightly better because for SVD the cut-off index $i$ as an integer has to be rounded. The reconstruction is rather bad because of the low SNR (high standard deviation in Fig. 13). Fig. 17 shows the reconstruction results for a reduced standard deviation ($10^{-5}$ in Fourier space, compared to 1 in Fig. 16). For smaller fluctuations (= standard deviation) SVD and Tikhonov reconstruction get similar because the effect of rounding to an integer for truncated SVD gets smaller if the cut-off-index gets higher.

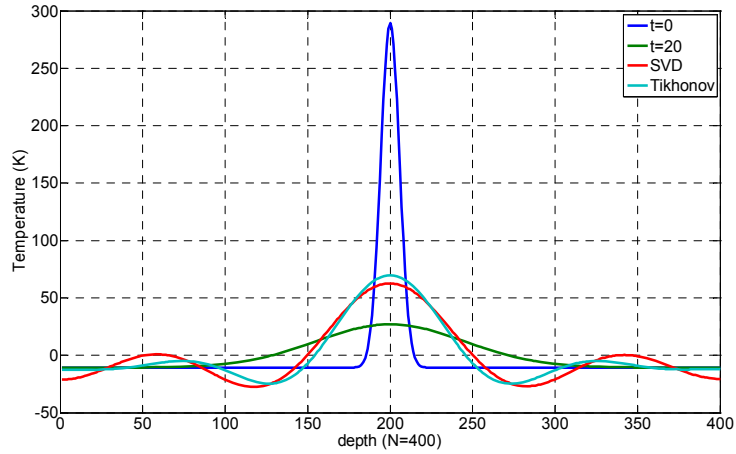

*Fig. 16. Reconstruction using the time evolution factors from Fig. 15 for SVD and Tikhonov reconstruction.*

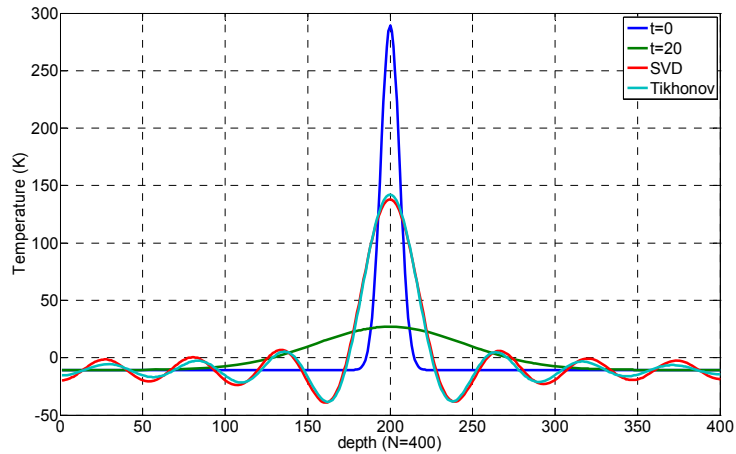

*Fig. 17. Reconstruction results for a reduced standard deviation ($10^{-5}$ in Fourier space, compared to 1 in Fig. 16).*

The width of the reconstructed peak is smaller for the higher SNR and therefore enables a better spatial resolution. The spatial resolution is defined as the minimal distance of two peaks, which can still be reconstructed as two individual peaks. This is equal to the width of the peak if the initial temperature peak at $t = 0$ is a sharp peak like a delta-source in space and time. The analytic expression is shown in eq. (41) and shows a width (standard deviation) of $\sqrt{2\alpha t}$. In k-space $\hat{T}(k,t) = \exp(-k^2 \alpha t)$ and the width is the inverse $\frac{1}{\sqrt{2\alpha t}}$.

In k-space SVD reconstruction (eq. (49)) and in a good approximation for higher signal-to-noise ratios (SNRs) also Thikhonov reconstruction (eq. (51)) gives a rectangular function up to $k_i$ with $\exp(k_i^2 \alpha t) = SNR$. After inverse Fourier transformation (eq. (41) with an integral from $-k_i$ to $+k_i$) this gives for the initial time $t = 0$ a *sinc* function in real space for the reconstructed profile (compare reconstruction results in Fig. 16 and Fig. 17)

$$(54) \qquad T_{reconstruction}(x, t=0) = \sqrt{\frac{2}{\pi}} \frac{\sin(k_i x)}{x} \quad with \quad k_i = \sqrt{\frac{\ln(SNR)}{\alpha t}}.$$

In Fig. 18 the spatial resolution (defined as the width between the two inflection points of the central peak of the *sinc* function) for a reconstructed initial temperature profile from the measured temperature after a certain time is shown. This width can be approximated for higher SNRs by the distance between the zero points of the reconstructed temperature profile:

$$(55) \qquad resolution = \frac{2\pi}{k_i} = 2\pi \sqrt{\alpha t / \ln(SNR)}.$$

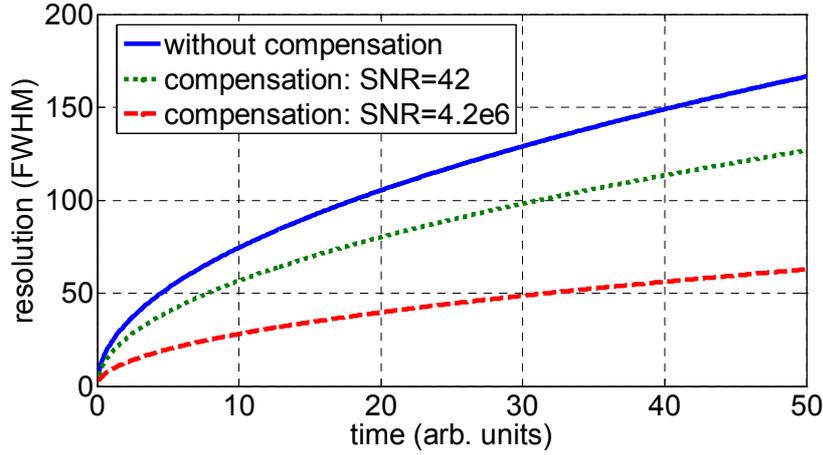

*Fig. 18. Possible spatial resolution (FWHM) from a measured temperature profile after a certain time without compensation (solid line) and with compensation for a small signal-to-noise-ratio (SNR) from the data shown in Fig. 13 (dotted line) and a SNR of $4.2 \cdot 10^6$ (dashed line).*

A further application of the presented theory is using the measured temperature profiles to determine the thermal diffusivity $\alpha$ as a material property. The accuracy of the determined thermal diffusivity from those measurements can be derived also by taking the fluctuations of the thermal waves into account.

### 3.4. Thermal diffusion and its inverse in ω –space

Up to now the whole 1D temperature distribution was measured at a certain time to reconstruct the initial temperature distribution. In non-destructive imaging, as sketched in Fig. 1, the signal is measured at the surface of the sample as a function of time and now the initial temperature distribution should be reconstructed from the measured surface temperature.

This situation can be described in "ω-space" (Fig. 11, left). The thermal waves have an angular frequency ω and $\sqrt{\omega/2\alpha}$ as wavenumber in one dimension (eq. (40)). The real part of $\theta$ is the cosine-transform of the temperature $T(x,t)$

(56)
$$\operatorname{Re}\theta(x,\omega) = \int_0^{+\infty} T(x,t)\cos(\omega t)dt = \frac{1}{2\sqrt{\alpha\omega}}\exp(-\sqrt{\frac{\omega}{2\alpha}}|x|)\cos(\sqrt{\frac{\omega}{2\alpha}}|x|+\frac{\pi}{4})$$
$$= \operatorname{Re}\theta(x=0,\omega)\exp(-\sqrt{\frac{\omega}{2\alpha}}|x|)[\cos(\sqrt{\frac{\omega}{2\alpha}}|x|)-\sin(\sqrt{\frac{\omega}{2\alpha}}|x|)]$$

In the section 3.2 we had a stochastic process (Ornstein-Uhlenbeck) to describe the time evolution of the Fourier series coefficients in "k-space" (Fig. 12). Now the Fourier coefficients for $\operatorname{Re}\theta(\omega,x)$ will be described in "ω-space" by the same stochastic process as the damped harmonic oscillator in chapter 2.1.3:

(57)
$$\frac{dD(x)}{dx} = \operatorname{Re}\theta(x)$$

(58)
$$\frac{d\operatorname{Re}\theta(x)}{dx} = -2\sqrt{\frac{\omega}{2\alpha}}\operatorname{Re}\theta(t) - \frac{\omega}{\alpha}D + \sigma\,\eta(t).$$

At a distance $x$ the amplitude is damped by $\exp(-x\sqrt{\omega/2\alpha})$. Fig. 19 shows $\operatorname{Re}\theta$ as a function of the scaled distance $x\sqrt{\omega/\alpha}$ as a realization of a stochastic process and the mean value as described in eq. (56). $\operatorname{Re}\theta$ is scaled that the standard deviation is one.

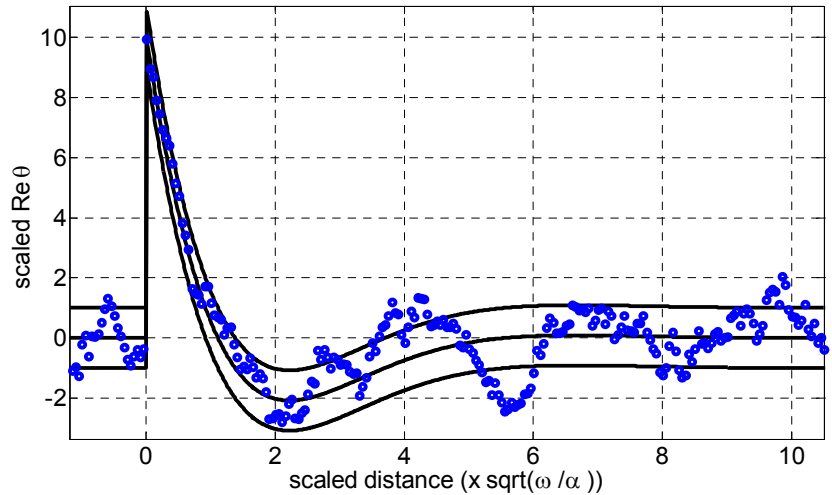

*Fig. 19. Points show a realization of the stochastically damped oscillator; the solid lines represent the mean and the standard deviation (see also Fig. 9). The mean value is described by eq. (56). $\operatorname{Re}\theta$ is scaled that its standard deviation is one and the scaled distance is $x\sqrt{\omega/\alpha}$.*

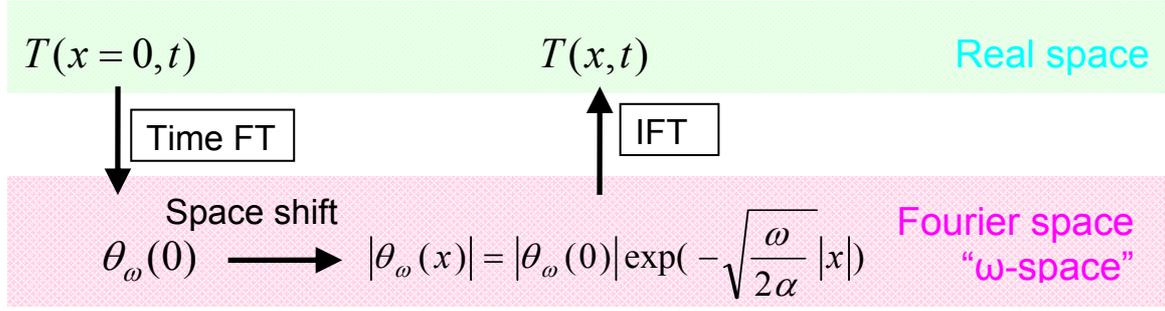

*Fig. 20: The temperature T at x=0 is time Fourier transformed (Time FT). The space shifted Fourier series coefficients can be described similar to the mean value of a stochastically damped oscillator process (eq.(28)). The temperature at x is then calculated by an inverse Fourier transform (IFT).*

Analog to Fig. 12, where the time shifted temperature distribution is calculated, in Fig. 20 the procedure to calculate a space shifted temperature evolution is shown. As an example we use the Gaussian peak from Fig. 13 as initial temperature, but at a depth of 20. Then the surface temperature can be calculated as the inverse cosine transform of $\text{Re}\,\theta$ from eq. (56). The deviation of the surface temperature from the equilibrium temperature described by this stochastic process is shown in Fig. 21. The standard deviation in real space is $s_k \sqrt{(N-1)/2}$ (the same as in Fig. 13) and the standard deviation in ω-space is $s_k \sqrt{(N-1)/(N_t-1)}$, which is the reciprocal value of *SNR*. In the shown example in real space *N=400* and the number of time points was chosen $N_t=1000$ at a *Δt=0.2*.

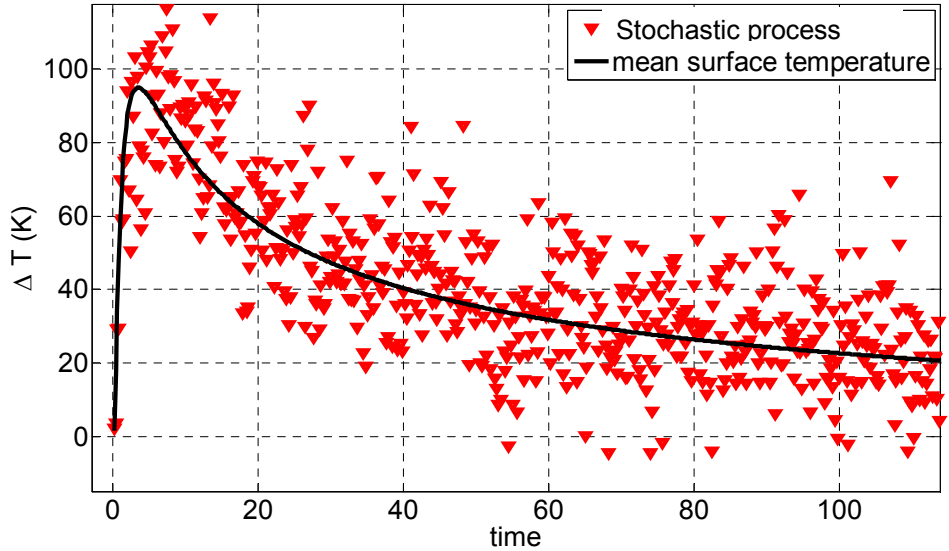

*Fig. 21. Deviation from equilibrium temperature $\Delta T = T - \bar{T}$ at the surface as a function of time, "measured" at a time step of Δt = 0.2. The initial temperature distribution is Gaussian shaped with a maximum of 600 Kelvin (K) at a depth of 20 and the full width at 1/e of the maximum is 16 arb. units (compare Fig. 13). The minimum temperature is 300 K. The standard deviation at the surface is again constant in time.*

For the reconstruction of the initial temperature profile from a "measured" surface temperature like in Fig. 21 we can use a cut-off frequency similar to the cut-off wave vector (for comparison see Fig. 15). For a certain distance *a* the amplitude is decreased by $\exp(-a\sqrt{\omega/2\alpha})$, which can be set equal to the noise level in ω-space. Therefore the cut-off frequency is:

*(59)* $$\omega_i = 2\alpha(\ln(SNR)/a)^2.$$

Here *SNR* is the signal-to-noise ratio of the time Fourier transformed surface temperature signal in ω-space:
$$SNR = SNR \text{ in real space } \sqrt{(N_t-1)/2}$$

For a delta-source in one dimension at a distance *a* the temperature is given by eq. (40). Using the cut-off frequency $\omega_i$ for the upper integration limit one gets for the initial temperature at *t = 0* (integral can be solved analytically):

*(60)*
$$T(x,t=0) = \frac{1}{\pi}\int_0^{\omega_i} \cos\left(\sqrt{\frac{\omega}{2\alpha}}|x-a|+\frac{\pi}{4}\right)\frac{1}{\sqrt{\omega\alpha}}\exp\left(-\sqrt{\frac{\omega}{2\alpha}}|x-a|\right)d\omega = \frac{2}{\pi}\frac{\sin(k(x-a))}{x-a}\exp(-k|x-a|)$$
$$\text{with } k = \ln(SNR)/a$$

Fig. 22 shows the reconstructed temperature for a delta-source at a depth of 20 and 40. For each of the two depths three different noise levels have been compared: $s_k^2 = 10^{-5}, 10^{-4}, \text{ and } 10^{-3}$ is the variance in Fourier space (see section 3.2).

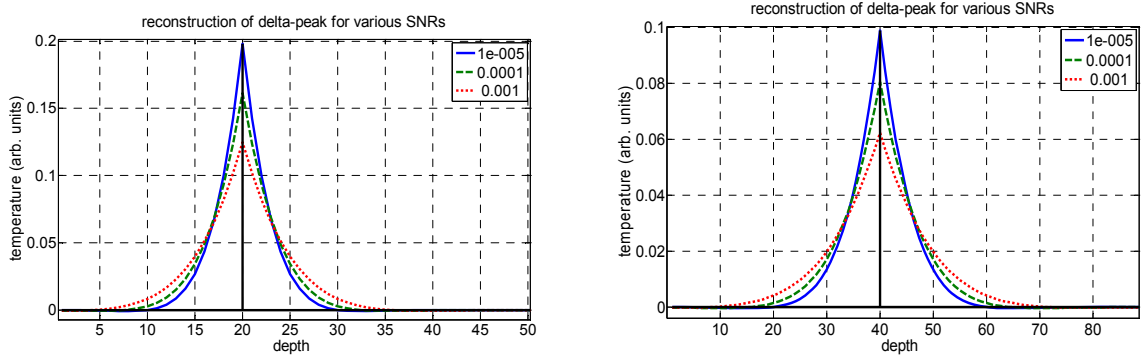

*Fig. 22. Reconstruction of the temperature profile from a delta-source at a depth of 20 (left) and 40 (right) for three different noise levels.*

The resolution is the smallest distance at which two different peaks can be resolved in the reconstructed temperature and can be estimated by the distance of the two zero points near the peak at *x = a*:

*(61)* $$resolution = 2\pi\, a / \ln(SNR)$$

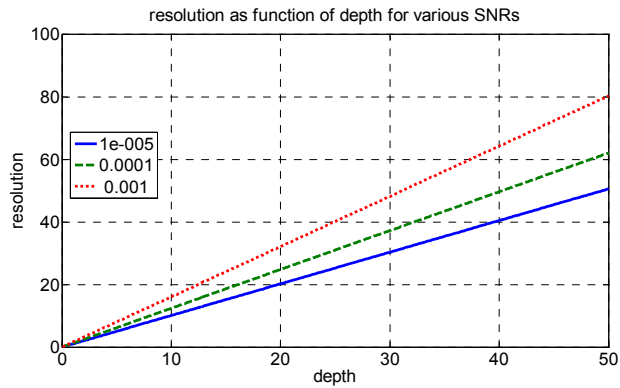

*Fig. 23. Resolution as function of depth for various SNRs. The resolution is the smallest distance at which two different peaks can be resolved in the reconstructed temperature and is estimated by the distance of the two zero points near the peak at the depth (see also Fig. 22).*

Fig. 23 shows the resolution as a function of depth for the three noise levels. *For reconstructions from the measured surface temperature as a function of time it is proportional to depth and indirect proportional to the natural logarithm of the signal-to-noise ratio.* This is the main result of this section. For reconstructions from the measured temperature on the whole sample at a certain time the resolution is proportional to the square root of time and indirect proportional to the square root of the natural logarithm of the signal-to-noise ratio (see Fig. 18 in the last section).

### 3.5. Numerical example to compare reconstruction in k-space and ω-space

In this section the resolution derived by using the cut-off frequency $\omega_i$ (ω-space in the left of Fig. 11) is compared with the results from *k*-space (right "wing" of Fig. 11). As an example we use the Gaussian peak (compare Fig. 13) as initial temperature at a depth of 20 and 40, and a full width at 1/e of the maximum of 4 arb. units (to simulate a delta-source). For each of the two depths a noise level of $s_k^2 = 10^{-3}$ has been used and the temperature as a function of time has been calculated up to a time of 200, like in Fig. 13, using a time step of *Δt=0.2 ($N_t$=1000)*. Then the surface temperature $T_{surface}$ at a depth $x = 0$ was recorded. Inserting eq. (43) in eq. (42) gives a relation between the temperature in real space as a function of time and the Fourier series coefficients. As this relation is valid on the whole sample space it is also valid on the surface:

$$(62) \qquad T_{surface}(t) = \sum_{k=0}^{N-1} \hat{T}_k(0) \exp(-\gamma_k t) \phi_k(x=0) = \sum_{k=0}^{N-1} \hat{T}_k(0) \exp(-\gamma_k t) .$$

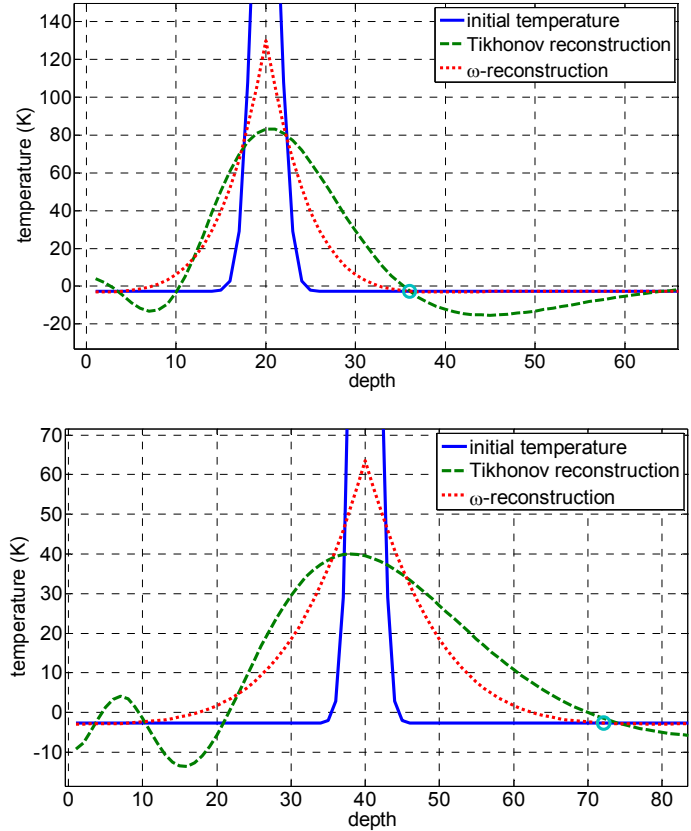

*Fig. 24. Comparison of the deviation from equilibrium temperature for Tikhonov reconstruction and "ω-space"- reconstruction for a Gaussian initial temperature peak centered at a depth of 20 (top) and at a depth of 40 (bottom). The zero point of the "ω-space"- reconstruction, which is the resolution described by eq. (61), is marked by a small circle 'o'.*

The last equality follows from $\phi_k = 1$ at the surface ($x = 0$) and by using the discrete time steps ($\Delta t=0.2$, $N_t=1000$) when the surface temperature was recorded we get a matrix equation from eq. (62). Inverting this matrix equation and calculating the Fourier coefficients $\hat{T}_k(0)$ from the surface temperature $T_{surface}$ is ill-posed. We used Tikhonov's regularization scheme[40] (see also section 3.3) and the regularization parameter was estimated by the "L-curve-method"[31]. The reconstructed initial temperature calculated by inverse Fourier transform is shown in Fig. 24. In this figure also the reconstructed temperature profile using the ω-space-reconstruction is shown and the zero point, which is the resolution described by eq. (61), is marked by a small circle 'o'. The resolutions for both reconstruction methods are similar. *This confirms for the used example of a Gaussian - shaped initial temperature profile at different depths the validity of eq. (61) to estimate the possible resolution for the reconstructed profile.*

## 4. Attenuated acoustic wave as a stochastic process

First results see section 5 of our Intech-chapter[24].

## 5. Summary, conclusions and outlook on future work

Future work is planned on random walk and correlations (compare eq. (9)) and on the question of causality for a random diffusion process (see e. g. Buckingham[37]). The method developed by La Riviere et al.[41] to compensate acoustic attenuation can be transferred to diffusive processes to enable efficient image reconstruction for thermography. Finally the boundary conditions should be changed from adiabatic boundary conditions to more realistic boundary condition of third kind. This more realistic simulation results can be compared with experimental data.

## 6. Acknowledgments

This work has been supported by the Christian Doppler Research Association, by the Federal Ministry of Economy, Family and Youth, by the Austrian Science Fund (FWF) project numbers S10503-N20 and TRP102-N20, by the European Regional Development Fund (EFRE) in the framework of the EU-program Regio 13, and the federal state Upper Austria.